\documentclass[useAMS,usenatbib]{mn2e}

\usepackage{graphicx}

\pdfoutput=1

\def\lesssim{\mathrel{\hbox{\rlap{\hbox{\lower4pt\hbox{$\sim$}}}\hbox{$<$}}}} 
\def\gtrsim{\mathrel{\hbox{\rlap{\hbox{\lower4pt\hbox{$\sim$}}}\hbox{$>$}}}} 

\DeclareRobustCommand{\ion}[2]{%
\relax\ifmmode
\ifx\testbx\f@series
{\mathbf{#1\,\mathsc{#2}}}\else
{\mathrm{#1\,\mathsc{#2}}}\fi
\else\textup{#1\,{\mdseries\textsc{#2}}}%
\fi}

\newcommand{\HII}{\mbox{H\,{\sc ii}}}
\newcommand{\NIIs}{\mbox{\scriptsize N\,{\sc ii}}}

\title[Explosion Sites of GRBs \& CC SNe]{
Metallicity Measurements of Gamma-Ray Burst and Supernova Explosion Sites: 
Lessons from H\,{\Large\bf II} regions in M31}

\author[Y. Niino, K. Nagamine and B. Zhang]{Yuu Niino$^{1}$\thanks{E-mail: yuu.niino@nao.ac.jp}, 
Kentaro Nagamine$^{2,3}$ \& Bing Zhang$^{3}$\\
 $^{1}$Division of Optical \& Infrared Astronomy, National Astronomical Observatory of Japan, 
2-21-1 Osawa, Mitaka, Tokyo 181-8588, Japan\\
 $^{2}$Department of Earth and Space Science, Graduate School of Science, Osaka University, \\
1-1 Machikaneyama-cho, Toyonaka, Osaka, 560-0043, Japan\\
 $^{3}$Department of Physics and Astronomy, University of Nevada, Las Vegas, 
4505 S. Maryland Pkwy, Las Vegas, NV 89154-4002, U.S.A.}

\begin{document}

\date{}

\pagerange{\pageref{firstpage}--\pageref{lastpage}} \pubyear{2014}

\maketitle

\label{firstpage}

\begin{abstract}
We examine how the small-scale ($<$\,kpc) variation of metallicity 
within a galaxy, which is found in nearby galaxies, affect 
the observational estimates of metallicity in the explosion sites of transient events 
such as core-collapse supernovae (CC SNe) and gamma-ray bursts (GRBs). 
Assuming the same luminosity, metallicity, 
and spatial distributions of \ion{H}{ii}\ regions (hereafter HIIRs) as observed in M31, 
we compute the apparent metallicities that we would obtain 
when the spectrum of a target region is blended with those of surrounding HIIRs
within the length scale of typical spatial resolution. 
When the spatial resolution of spectroscopy is $\lesssim$ 0.5\,kpc, 
which is typical for the existing studies of CC SN sites, 
we find that the apparent metallicities 
reflect the metallicities of target regions, but with 
significant systematic uncertainties in some cases. 
When the spatial resolution is $\gtrsim$ 1.0 kpc,
regardless of the target regions 
(which has a wide range of metallicity that spans $\sim$ 0.6 dex for the M31 HIIRs), 
we always obtain the apparent metallicities similar to the average metallicity of the M31 HIIRs. 
Given that the apparent metallicities measured 
with $\gtrsim$\,kpc scale resolution do not necessarily reflect 
the immediate environment of the stellar explosions, 
the current observational estimates of high metallicities for some of the long GRB host galaxies 
do {\em not} rule out the hypothesis that the long GRBs 
are exclusively born in a low-metallicity environment. 

\end{abstract}

\begin{keywords}
gamma-ray burst: general -- supernovae: general -- galaxies: abundances -- ISM: H II regions.
\end{keywords}

\section{Introduction}

The nature of the progenitors of stellar explosions 
is one of the most important questions in astronomy. 
It is generally agreed that type II, Ib, and Ic supernovae 
(core collapse supernovae, hereafter CC SNe) 
and most long-duration gamma-ray bursts (long GRBs) 
originate from core collapses of massive stars 
with $\gtrsim 8M_\odot$ at the end of their lives. 
Despite the above theoretical framework, 
the very physical reasons that define diverse supernova types 
and possible GRB associations with core collapse events are not clearly identified. 

Some theoretical studies on the origin of GRBs using stellar evolution models suggest 
that a low metallicity may be a necessary condition for a GRB to occur 
\citep[$Z < $ a few $\times\ 0.1Z_\odot$; e.g.,][]{Yoon:2005a,Yoon:2006a,Woosley:2006a}. 
Observational studies have also shown 
that the metallicity distribution of the GRB host galaxies at redshifts $z \lesssim 1$ 
is significantly biased towards lower metallicities than 
that of general late-type galaxies at similar redshifts \citep{Stanek:2006a, Graham:2013a}. 

However, the metallicity of a host galaxy is not necessarily identical 
to that of the progenitor of a stellar explosion that occurs in it. 
There might be systematic differences
between metallicities of the host galaxies and the progenitors. 
\citet{Niino:2011b} discussed the metallicity distribution 
of galaxies that harbour low-metallicity star formation 
considering the observed properties of the local galaxies 
including the internal variation of metallicity within each galaxy, 
and showed that up to $\sim$ 25 per cent 
of cosmic low-metallicity star formation with 12+log(O/H) $<$ 8.2 
takes place in high-metallicity galaxies with 12+log(O/H) $>$ 8.8. 
To address this issue, some observational studies 
tried to spatially resolve some GRB host galaxies 
and measure the metallicity of the local environment 
at the GRB sites \citep[e.g.,][]{Modjaz:2008a}. 
The systematic metallicity differences between the transient event sites 
and other parts of the host galaxies are actually observed in some cases 
\citep[e.g.,][]{Levesque:2011a, Sanders:2012b, Taddia:2013a}. 
For long GRB host galaxies, the site metallicity
tends to be lower than that in the other parts of galaxies. 
However, it is also claimed that the explosion sites 
of GRB 020819 \citep{Levesque:2010b} and GRB 120422A 
\citep[][but see also \citeauthor{Levesque:2012a}~\citeyear{Levesque:2012a}]{Schulze:2014a} have high metallicities. 

The explosion site metallicities of CC SNe 
\citep[e.g.,][]{Anderson:2010a, Modjaz:2011a, Galbany:2014a} are also systematically investigated. 
Among the major classes of CC SNe (II, Ib, and Ic), 
SNe Ib and Ic tend to occur in high-metallicity regions compared to SNe II \citep{Anderson:2010a}, 
although the difference of the site metallicities of type Ib and Ic SNe is still a matter of debate 
\citep{Modjaz:2011a, Leloudas:2011a, Sanders:2012b, Kuncarayakti:2013a}. 
It is also suggested that metallicity plays an important role 
in the occurrence of some subclasses of CC SNe \citep{Modjaz:2011a, Sanders:2012a, Taddia:2013a, Lunnan:2014a}. 

It should be noted that we are not always able to resolve 
individual \ion{H}{ii}\ regions (HIIRs) that hosted GRBs and/or CC SNe, 
and hence the obtained spectra of the explosion sites 
may be blended with other neighboring HIIRs. 
In the context of supernova remnant (SNR) searches, 
\citet{Matonick:1997a} have shown that 
the observed number density of SNR is higher in nearer galaxies, 
because it becomes easier to resolve a confused region in a galaxy when it is nearby. 
We can expect that our capability to investigate the explosion site 
of a transient event would depend on the distance to the event 
\citep[e.g.,][]{Sanders:2012b, Sanders:2013a, Taddia:2013a}, 
as in the cases of SNR searches. 

Recently, \citet{Sanders:2012b} discussed how the metallicity measurements 
of Type Ibc SNe sites would be affected by the spatial resolution, 
assuming that an intrinsic metallicity difference of 0.2 dex 
is unrecognizable due to blending when the spatial resolution is worse than 2\,kpc. 
Their results suggested that the blending does not 
significantly affect their SN sample whose typical redshift is $< 0.1$. 
However, it is not known what spatial resolution is really necessary  
to probe the immediate environment of a transient, 
which would be closely connected to the nature of the progenitor star. 
Especially for GRBs that typically occur at redshifts $\gtrsim$ a few $\times$ 0.1, 
the typical spatial resolution of ground-based observations ($\sim 1$ arcsec) 
corresponds to a few kpc, which is often limited by the seeing of atmosphere (not instrument).  
Therefore it is likely that there is some metallicity variation below the resolution limit. 

The internal metallicity structure of galaxies have been studied for decades. 
It is broadly agreed that the metallicity decreases as the galactocentric radius increases 
in many galaxies \citep[so-called metallicity gradient, e.g.,][]{Shields:1978a, Zaritsky:1994a}, 
while slope of the gradient may vary within a galaxy 
\citep[e.g.,][]{Luck:2003a, Bresolin:2009a, Balser:2011a, Esteban:2013a}, 
and a significant scatter of metallicity around the gradient may also exist 
\citep[e.g.,][]{Afflerbach:1997a, Rosolowsky:2008a}. 
Some recent studies intensely performed integral field spectroscopic (IFS) observations 
of nearby late type galaxies \citep[e.g.,][]{Rosales-Ortega:2010a, Marmol-Queralto:2011a, 
Sanchez:2012a, Fogarty:2012a, Blanc:2013a, Richards:2014a, Belfiore:2014a}, 
and dramatically increased the sample size of HIIRs with measured metallicities. 
Furthermore, \citet[][hereafter S12]{Sanders:2012c} obtained the spectra 
of $> 200$ HIIRs in M31 (the Andromeda galaxy), 
and found that $\sim 1/3$ of the HIIR pairs with separations less than 500 pc 
show significant (i.e., larger than the error) metallicity variation. 

In this paper, we examine how the metallicity estimates from 
spectroscopic observations are affected by limited resolution. 
We do this by performing mock blended observations with limited resolution, 
assuming the same observed distributions of emission-line luminosities and line ratios as the M31 HIIRs. 

The remaining part of the paper is organized as follows. 
In \S\ref{sec:data}, we describe the spectroscopic and photometric data sets that we use in this study. 
In \S\ref{sec:metal}, we discuss the variation of metallicity 
among the M31 HIIRs, especially on small scales ($\lesssim$\,kpc). 
In \S\ref{sec:lum}, we discuss the line luminosities of HIIRs. 
In \S\ref{sec:demo}, we demonstrate the degree of bias in the measured metallicity 
caused by the small-scale variation in metallicity. 
In \S\ref{sec:discussion}, we discuss the implications 
for the current observations of the explosion sites of transient events. 
We summarise our conclusions in \S\ref{sec:fin}. 

\section{Data sets}
\label{sec:data}

We use the observed properties of the HIIRs in M31 
to investigate how the spatial resolution limit affects 
the metallicity estimates of explosion sites. 
M31 is an irreplaceable laboratory to study 
the small-scale variation of metallicity in interstellar medium (ISM) of a late type galaxy.  
In M31, the structures of HIIRs are resolved 
down to $< 10$\,pc \citep[e.g.,][]{Massey:2007a, Azimlu:2011a}, 
and spectroscopic information is available for more than $200$ HIIRs (S12). 
The HIIRs in the Magellanic clouds have also been studied for decades, 
however, the number of spectroscopic sample is small for a statistical study 
\citep[$\sim$ 20; e.g.,][]{Dufour:1977a, Pagel:1978a, Vermeij:2002a}. 
Although the observed sample size of HIIRs in other local late-type galaxies 
was significantly increased by the recent IFS surveys, 
their spatial resolution was $\gtrsim 100$\,pc 
except for a few cases \citep{Sanchez:2012a, Sanchez:2014b}, 
and many HIIRs that have a spatial extent of $\lesssim$ several 10\,pc are not resolved. 

\citet[hereafter A11]{Azimlu:2011a} constructed a photometric sample of HIIRs 
based on broad- and narrow-band images taken by the Local Group Galaxies Survey 
\citep[LGGS,][]{Massey:2006a}, covering the whole disk of M31. 
The sample contains $3961$ HIIRs 
with H$\alpha$ luminosities $L_{\rm H\alpha} \gtrsim 10^{34.5}$ erg s$^{-1}$,  
excluding known and potential planetary nebulae (PNe). 
The largest spectroscopic sample of HIIRs in M31 was constructed by S12, 
who obtained the spectra of 253 HIIRs and 407 PNe, 
selected from the LGGS images and some samples of emission line objects in the literature. 

In this study, we use the photometric and spectroscopic sample 
of the M31 HIIRs provided by A11 and S12 to examine  
how the spatial resolution affects the observed properties of transient event sites. 
Following S12, we obtain the HIIR positions 
in a deprojected coordinate on the disk of M31
assuming the following quantities: 
the inclination angle of $12^\circ.5$ \citep{Simien:1978a};  
the distance to M31 of 770\,kpc \citep{Freedman:1990a};  
the position of M31 center ($\alpha$, $\delta$) 
and the angle of disk major axis relative to north celestial pole ($\phi$) as
\begin{eqnarray} 
\alpha &=& 00^\circ42'44''.52\ ({\rm J2000}), \\
\delta &=& +41^\circ16'08''.69\ ({\rm J2000}), \\
\phi &=& 37^\circ42'54'' 
\end{eqnarray} 
 \citep{Baade:1964a}.

\section{Metallicities of the \HII\ regions}
\label{sec:metal}

\subsection{Metallicity diagnostics}
\label{sec:diagnostic}

S12 obtained the fluxes of following emission lines:
 [\ion{O}{ii}]~$\lambda$3727, [\ion{O}{iii}]~$\lambda\lambda$4363,4959,5007, 
H$\beta$, [\ion{N}{ii}]~$\lambda\lambda$6548,6584, H$\alpha$, and [\ion{S}{ii}]~$\lambda\lambda$6717,6731. 
Not all of these lines are detected for all HIIRs in the S12 sample. 
It should be noted that S12 obtained the line fluxes with 1.5 arcsec fibers. 
The fiber loss corrections are not available in S12, 
thus the fluxes are different from the total fluxes. 
Therefore we only use the ratio between different lines in S12 sample, but not the absolute flux values. 

Various metallicity calibration methods are proposed 
to measure the metallicity of ionized gas in HIIRs.
However, the results of different calibration methods 
are not always consistent with each other \citep[e.g.,][]{Kennicutt:2003a, Kewley:2008a, Lopez-Sanchez:2012a}.  
In order to maximize our sample size with metallicity estimates, 
we mainly use the N2 index = log$_{10}$[\ion{N}{ii}]/H$\alpha$ as a metallicity indicator 
(hereafter [\ion{N}{ii}] means [\ion{N}{ii}]~$\lambda$6584 unless otherwise specified). 
Among the 253 HIIRs in the S12 sample, H$\alpha$ and [\ion{N}{ii}] lines 
are detected for 222 HIIRs with flux errors $< 20$ per cent, 
while [\ion{O}{iii}]~$\lambda$5007, which is also widely used to measure the HIIR metallicity, 
is detected only for 58 HIIRs with flux errors $< 20$ per cent. 
We also discuss the analysis with other metallicity indicators in \S\ref{sec:otherZ}. 
The metallicity indices are corrected for dust extinction 
using the extinction curve by \citet[][$R_V = 3.1$]{Cardelli:1989a} 
and $A_V$ of each HIIR, which were derived by S12 
assuming an intrinsic value of H$\alpha$/H$\beta = 2.85$ and Cardelli's extinction curve. 

We assume a relation between the N2 index and 12+log(O/H) 
empirically calibrated by \citet{Marino:2013a} using an observed sample of HIIRs in local galaxies 
with metallicity measurements by the `direct' method \citep[e.g.,][]{Garnett:1992a}. 
It should be noted that the N2 index and many other metallicity indicators 
are also affected by the physical conditions of gas other than metallicity, such as the ionization state 
\citep[e.g.,][]{McGaugh:1991a, Pilyugin:2000a, Kewley:2002a, Lopez-Sanchez:2011a}. 
\citet{Marino:2013a} calibrated the N2 index with the HIIRs in local galaxies 
which would have physical conditions close to those of the M31 HIIRs discussed here, 
compared to the case of the calibrations based on galaxy scale observations \citep[e.g.,][]{Nagao:2006a, Maiolino:2008a}. 

It is known that the N2 index saturates when metallicity is high, 
and hence cannot be used to measure the high metallicity in some HIIRs. 
In the calibration by \citet{Marino:2013a}, the N2 index saturates 
when 12+log(O/H) $\gtrsim 8.7$ (N2 $> -0.2$), 
and we cannot determine the metallicity when the N2 index is higher than this value. 
S12 excluded the HIIRs with undetermined metallicities from their sample. 
Instead, we include all HIIRs with H$\alpha$ and [\ion{N}{ii}] detections in our sample, 
assuming that the HIIRs with N2 $> -0.2$ have higher metallicities  
than those with lower N2 index [i.e., 12+log(O/H)$_{\rm N2} \gtrsim 8.7$].

\subsection{Metallicity variation in M31}
\label{sec:scatter}

\begin{figure}
\includegraphics[width=84mm]{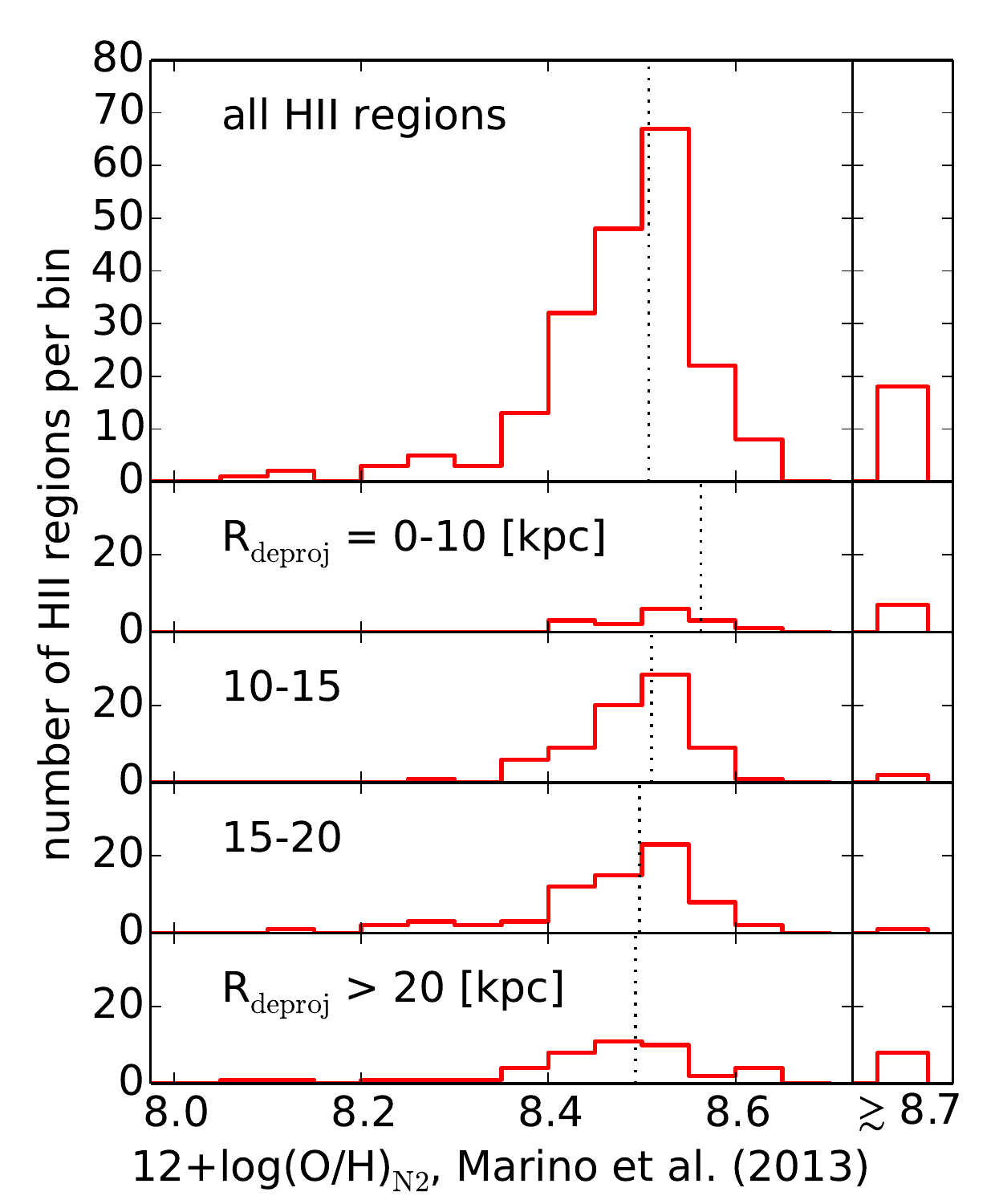}
\caption{
Metallicity distribution of the M31 HIIRs in the S12 sample, measured 
with the N2 index calibrated by \citet{Marino:2013a}. 
The top panel shows the distribution of all M31 HIIRs, 
and the other panels show the distributions 
at deprojected galactocentric radii $R_{\rm deproj} =$ 0--10, 10--15, 15--20 and $>$ 20\,kpc. 
The number of HIIRs with N2 $> -0.2$ (i.e., 12+log(O/H)$_{\rm N2} \gtrsim 8.7$) 
is shown separately in the right-hand-side of each panel. 
The vertical dotted lines represent the median of each distributions
including the HIIRs with N2 $> -0.2$. 
}
\label{fig:metaldist}
\end{figure}

We show the metallicity distribution of all M31 HIIRs in the top panel of Figure~\ref{fig:metaldist}. 
In the bottom 4 panels of Figure~\ref{fig:metaldist}, 
the sample is divided into 4 subsamples according to 
the deprojected galactocentric radius ($R_{\rm deproj}$). 
The HIIR metallicities at larger $R_{\rm deproj}$ are systematically lower. 
The large intrinsic scatter of metallicity discussed in S12 is clearly seen in each $R_{\rm deproj}$ bin. 
\citet{Marino:2013a} showed that the intrinsic error of their N2 index calibration $\pm$0.16 dex 
(or $\pm$0.09 dex depending on the calibration data sets).
Although the scatter may be partly due to the intrinsic error in the metallicity calibration methods, 
S12 showed that the HIIR-metallicity distributions in M31 have similarly large scatters
when they are measured by various calibration methods, 
some of which have the intrinsic errors $\lesssim$ 0.1 dex \citep{Kewley:2008a}, 
We also discuss other metallicity calibration methods in \S\ref{sec:otherZ}. 

\begin{figure}
\includegraphics[width=84mm]{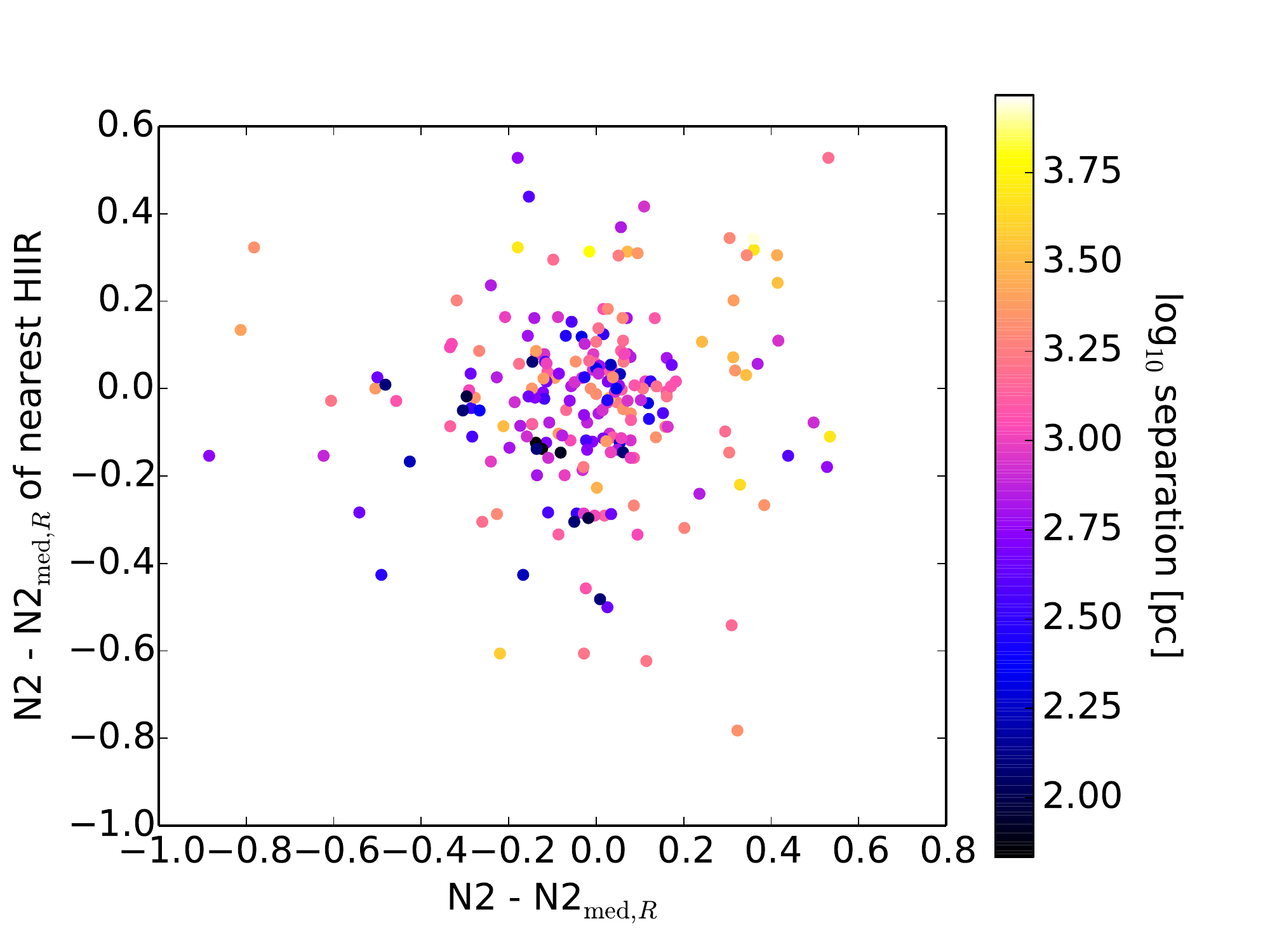}
\caption{
Comparison of the N2 index deviation of each HIIR 
from the median N2 in each radial bin (N2$-$N2$_{{\rm med},R}$) 
and that of the nearest one in the S12 sample. 
Separations of the HIIR pairs on the disk of M31 are colour-coded 
from dark blue (close, $\sim 100$\,pc) to bright yellow (distant, $\sim 8$\,kpc).  
}
\label{fig:pairLR}
\end{figure}

S12 pointed out that 33 per cent of the close HIIR pairs 
(with deprojected separations $<$ 0.5\,kpc)  
show metallicity variation of more than 0.3 dex. 
To further investigate the metallicity variation on small scales, 
we compare the N2 index of each HIIR and the nearest one for which N2 index is available. 
We divide the S12 sample into four different bins of $R_{\rm deproj}$ 
(0--10, 10--15, 15--20, and $> 20$\,kpc, as shown in Figure~\ref{fig:metaldist}), 
and consider the deviation of N2 index of each HIIR 
from the median N2 in each radial bin (N2$-$N2$_{{\rm med},R}$) 
to investigate the metallicity scatter separately from the metallicity gradient. 
In Figure~\ref{fig:pairLR}, we plot N2$-$N2$_{{\rm med},R}$ of each HIIR 
and N2$-$N2$_{{\rm med},R}$ of the nearest one to the HIIR. 
We do not find any significant correlation 
between N2$-$N2$_{{\rm med},R}$ of the neighboring HIIR pairs
including the cases with the deprojected separation of a few 100 pc. 
This suggests that the ISM in M31 is not mixed efficiently, 
and that the ISM metallicity varies even on small scales of a few 100 pc.  

\section{Line Luminosities of the \HII\ regions}
\label{sec:lum}

The emission-line luminosity distribution of HIIRs
is also an important issue when we discuss 
the blending of HIIRs in spectroscopy with limited spatial resolution. 
When we spectroscopically observe multiple HIIRs blended within spatial resolution, 
the one with stronger line emissions affects the resulting spectrum more,  
although the less luminous HIIRs in an emission line might be bright in other lines  
depending on their physical conditions such as metallicity and/or ionization state. 

To obtain the emission line luminosities of S12 HIIRs, 
we match the S12 sample to the photometric sample of A11, 
because the fiber-corrected total luminosity is not available for the S12 sample. 
We match each S12 HIIR to an A11 HIIR individually 
when the separation between the fiber position of the S12 HIIR 
and the flux peak location of the A11 HIIR 
is smaller than the HIIR radius determined by A11, 
and the S12 HIIR of concern is the nearest HIIR 
in the S12 sample to the A11 HIIR. 
It should be noted that, when the H$\alpha$ emission 
is extended with multiple intensity peaks, 
the HIIR detection method used in A11 
separates the object into multiple HIIRs. 
In such a case, it is possible that multiple A11 HIIRs 
match with a single S12 HIIR, 
in which case we assign the sum of the A11 HIIR luminosities
to the corresponding S12 HIIR. 
Among the 222 HIIRs with known N2 index in the S12 sample, 
197 have one or more counterparts in the A11 sample.  

To find the line emitting regions in M31, A11 used the H$\alpha$ narrow-band images 
with continuum subtraction based on the $R$-band images. 
The narrow-band contains H$\alpha$ and [\ion{N}{ii}] lines, 
and A11 corrected the [\ion{N}{ii}] line contamination to obtain the H$\alpha$ flux
assuming a global line ratio of [\ion{N}{ii}]/H$\alpha$ = 0.35. 
We undo this [\ion{N}{ii}] correction and recalculate the H$\alpha$ and [\ion{N}{ii}] fluxes 
based on the N2 index of each HIIR obtained by S12. 

\begin{figure}
\includegraphics[width=84mm]{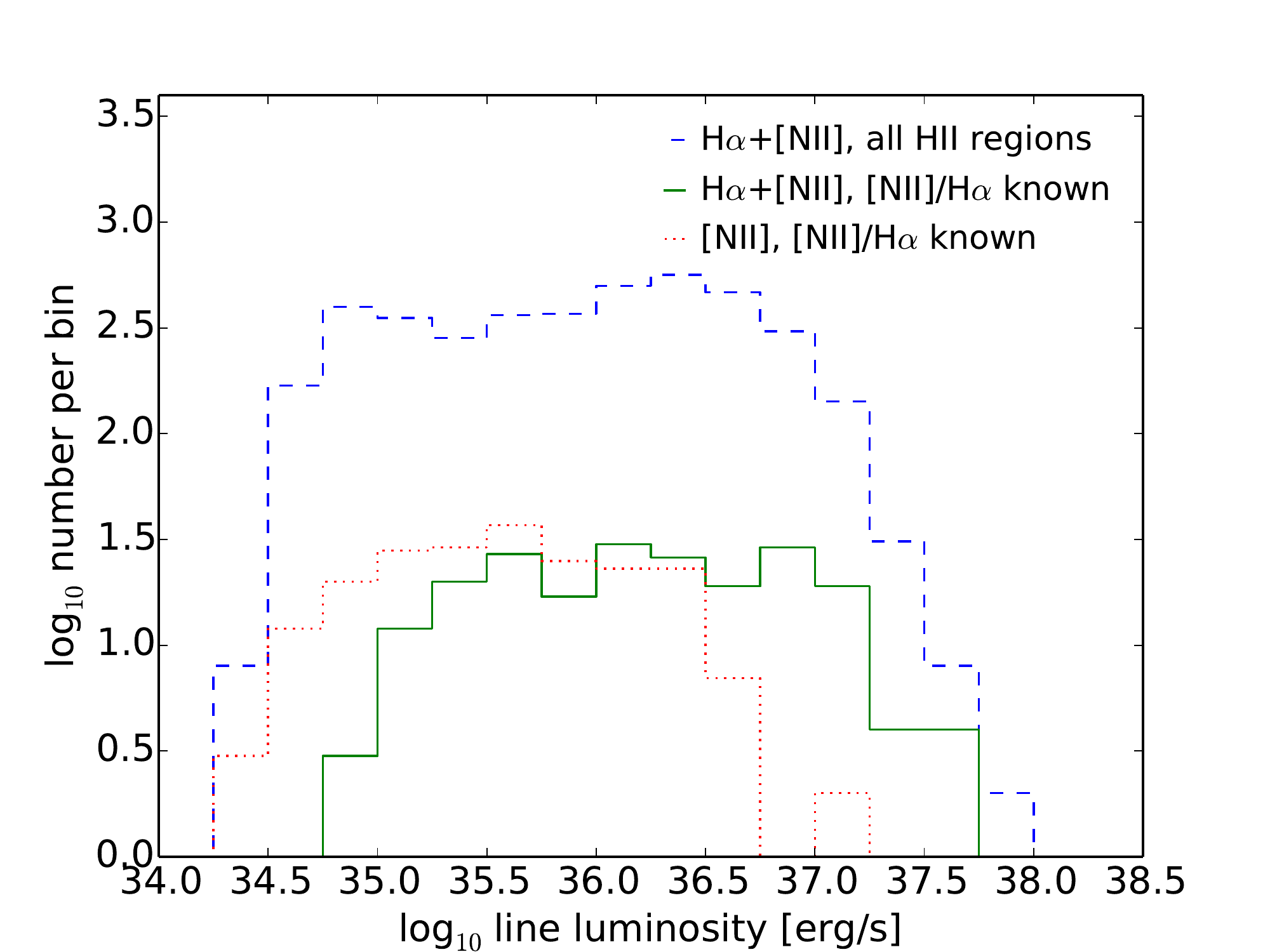}
\caption{
H$\alpha$ and [\ion{N}{ii}] emission line luminosity distributions 
of M31 HIIRs without dust correction. 
The dashed line shows the H$\alpha$+[\ion{N}{ii}] flux distribution 
of all A11 HIIRs including those without known N2 index. 
For those with known N2 index in the S11 sample, 
the distributions of H$\alpha$+[\ion{N}{ii}] flux (solid) and [\ion{N}{ii}] flux (dotted) are plotted. 
The line luminosities of S11 HIIRs are obtained by matching 
them to the A11 sample as described in the main text (\S\ref{sec:lum}). 
}
\label{fig:Lhadist}
\end{figure}

Figure~\ref{fig:Lhadist} compares the H$\alpha$+[\ion{N}{ii}] luminosity distribution 
of the S12 HIIRs with H$\alpha$ and [\ion{N}{ii}] detections (solid line)
to that of all A11 HIIRs (dashed line). 
Although the spectroscopic observations are performed 
only for a small fraction of the detected HIIRs, 
the sample with H$\alpha$ and [\ion{N}{ii}] detections covers 
a wide range of line luminosity, $35.0 < \log_{10} L_{\rm H\alpha+[\NIIs]} {\rm [erg\ s^{-1}]} < 37.5$, 
while [\ion{N}{ii}] is hardly observed for HIIRs 
with $\log_{10} L_{\rm H\alpha+[\NIIs]}  {\rm [erg\ s^{-1}]} < 35.0$. 
We also plot the [\ion{N}{ii}] luminosity distribution for the sample 
with H$\alpha$ and [\ion{N}{ii}] detections (dotted line). 
The faint end of the [\ion{N}{ii}] luminosity distribution 
suggests that the effective limiting luminosity for a line detection 
in the S12 spectroscopy was $\log_{10} L {\rm [erg\ s^{-1}]} \sim 34.5$, 
which is close to the limiting luminosity of the A11 sample. 
The [\ion{N}{ii}] luminosity distribution is $\sim$ 0.5 dex fainter than 
the H$\alpha$+[\ion{N}{ii}] luminosity distribution of the same sample, 
reflecting the typical N2 index $\sim -0.5$ in M31. 

Now let us consider a case where one tries to obtain the 
emission line ratio of a specific HIIR of interest 
(such as an explosion site of a GRB/CC SN) 
without sufficient spatial resolution to separate 
the HIIR from the surrounding ones. 
If the HIIR of interest is bright enough 
to dominate the total flux within the spatial resolution, 
we obtain the metallicity of the HIIR regardless of 
the metallicity distribution of surrounding HIIRs.
When we observe the explosion site of a long GRB and/or a SN Ic, 
the HIIR that hosted the transient is expected to be brighter than the other HIIRs,
because the progenitors of these explosions are likely very young 
\citep{Fruchter:2006a, Kelly:2008a, Leloudas:2010a}. 

\begin{figure}
\includegraphics[width=84mm]{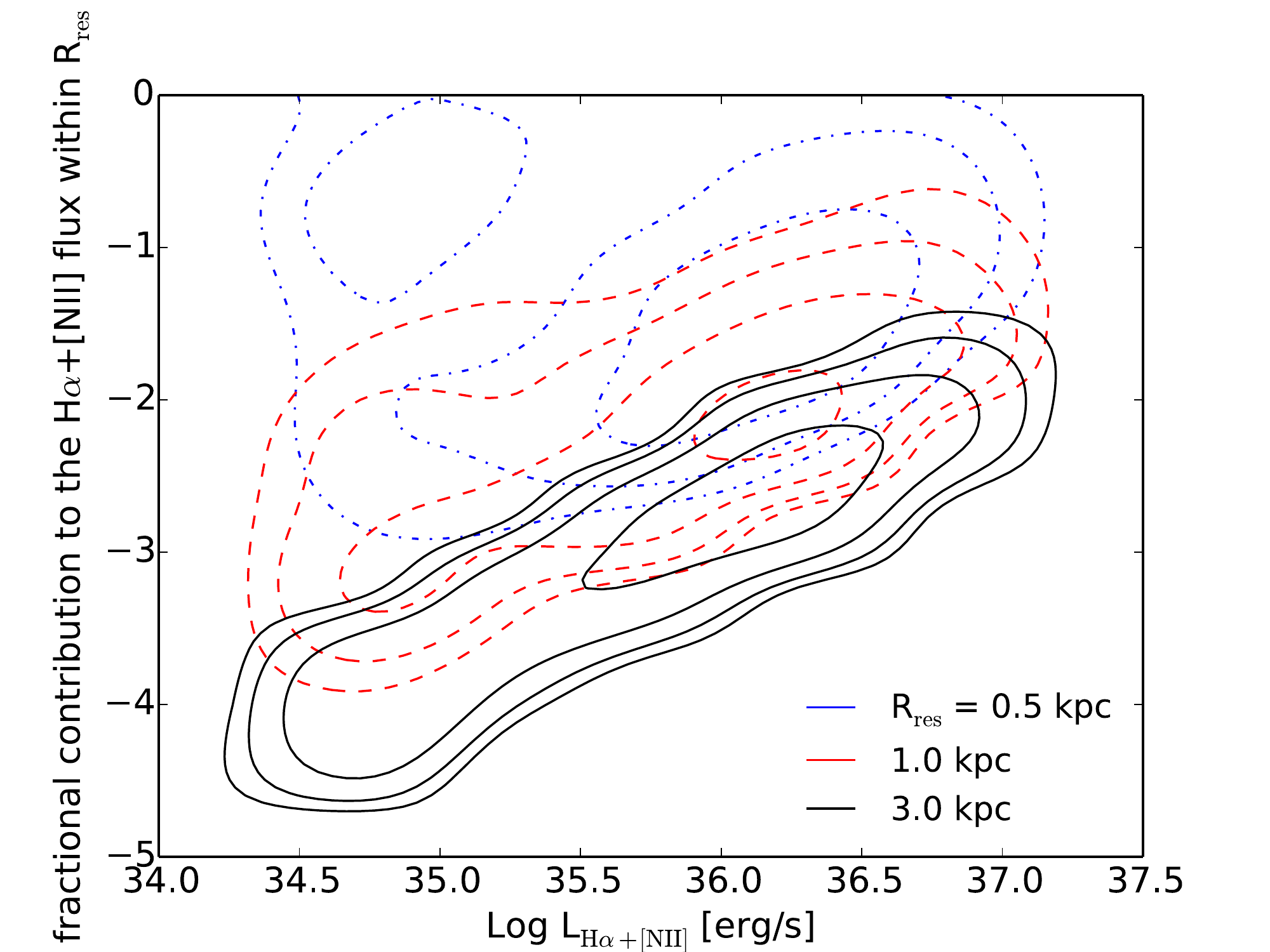}
\caption{
Distributions of the A11 HIIRs
on the $L_{\rm H\alpha+[\NIIs]}$ vs. fractional contribution plane 
with $R_{\rm res} =$ 0.5, 1.0, and 3.0\,kpc 
(dot-dashed, dashed, and solid, respectively). 
The contours indicate the surface number density 
of 100, 200, 400, and 800 dex$^{-2}$. 
}
\label{fig:fluxfrac}
\end{figure}

Using the photometric sample of 3961 HIIRs by A11, 
we investigate the fractional contribution of each HIIR
to the total H$\alpha$ luminosity 
within the resolution scale radius $R_{\rm res}$ in the deprojected disk coordinate. 
Here, the fractional contribution 
means the fraction of $L_{\rm H\alpha+[\NIIs]}$ 
to the total H$\alpha$+[\ion{N}{ii}] luminosity of all HIIR within $R_{\rm res}$. 
Figure~\ref{fig:fluxfrac} shows the fractional contributions of the A11 HIIRs 
versus $L_{\rm H\alpha}$ for different $R_{\rm res}$. 
For $R_{\rm res} \gtrsim 1.0$\,kpc, 
even the most brightest HIIR have 
a typical fractional contribution of $\lesssim$ 10 per cent, 
suggesting a significant blending effect on emission-line measurements. 

\begin{figure}
\includegraphics[width=84mm]{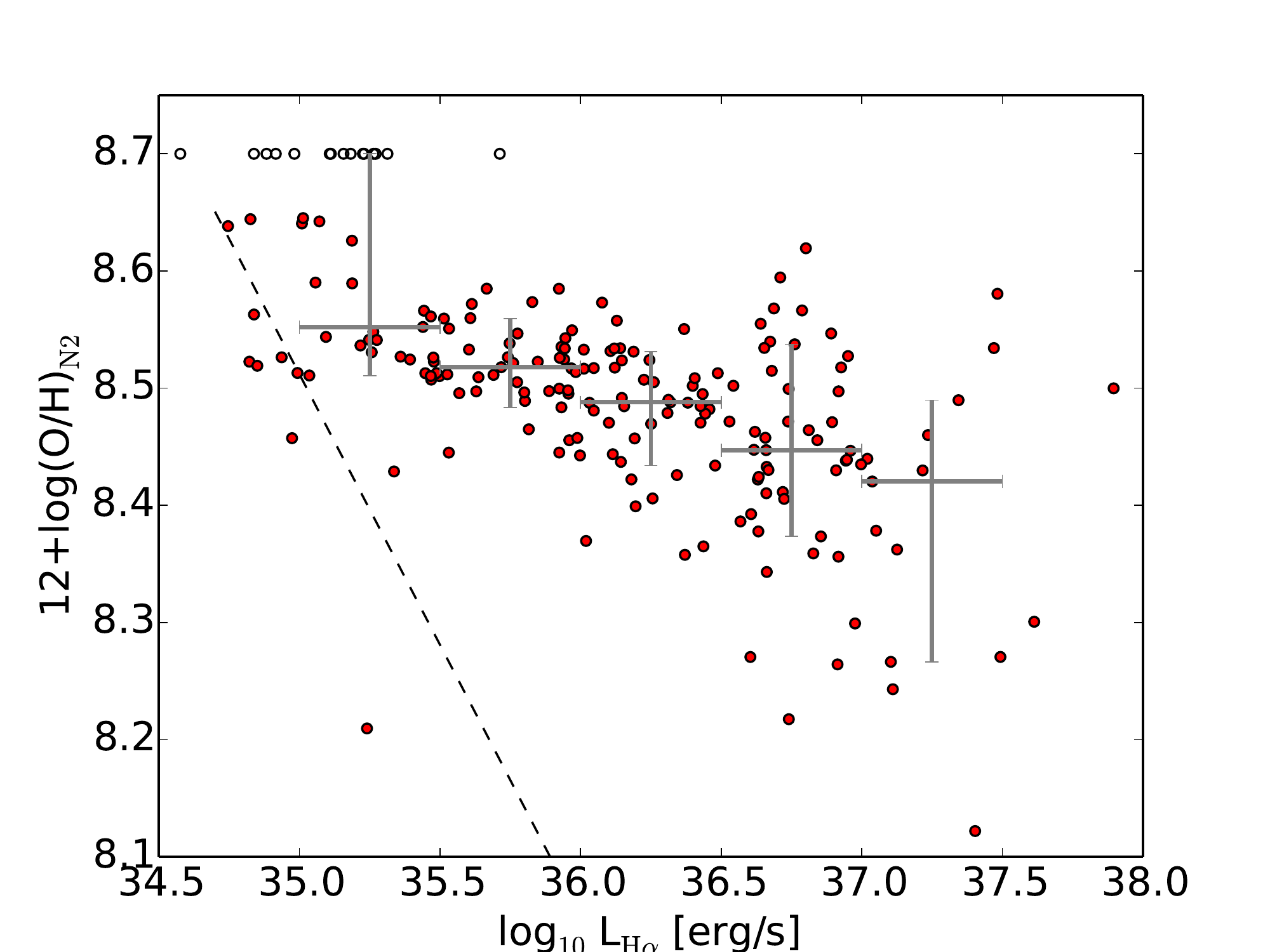}
\caption{
H$\alpha$ luminosities and metallicities of the HIIRs with known N2 index. 
The open circles at the top represent the HIIRs
with N2 $> -0.2$, for which the metallicities are not determined well. 
The dashed line represents the constant [\ion{N}{ii}] luminosity of $10^{34.5}$ erg s$^{-1}$. 
The error bars indicate the median and 68 percentile scatter 
in the five bins of $L_{\rm H\alpha}$ with the log-scale width of 0.5 dex 
in the range of log$_{10} L_{\rm H\alpha}$ [erg s$^{-1}$] = 35.0--37.5. 
}
\label{fig:LhaZ}
\end{figure}

Figure~\ref{fig:LhaZ} plots $L_{\rm H\alpha}$ versus metallicity of HIIRs with known N2 indices. 
It is clear that the brighter HIIRs tend to have lower metallicities. 
One possible cause of this trend is the non-detection of [\ion{N}{ii}] lines for the faint HIIRs. 
The N2 index positively correlates with metallicity, 
therefore it would be difficult to detect the [\ion{N}{ii}] line of a faint, metal-poor HIIR. 
The dashed line in Figure~\ref{fig:LhaZ} shows the relation for a fixed [\ion{N}{ii}] luminosity of $10^{34.5}$ erg s$^{-1}$. 
At the lowest $L_{\rm H\alpha}$, high-metallicity HIIRs naively follow this relation, 
but the low-metallicity HIIRs typically have much larger $L_{\rm H\alpha}$. 
Hence it might be difficult to explain the $L_{\rm H\alpha}$--metallicity relation 
only by a single limiting flux.
We note that the line detection limit 
may not be uniform over the entire M31 disk, as it is affected 
by the variation of the background radiation from diffuse interstellar gas (A11), 
and there might be PN contaminations in faint, high-metallicity HIIRs (S12). 

\section{Metallicity measurements with limited spatial resolution}
\label{sec:demo}

The small-scale metallicity variation and the luminosity distribution of HIIRs
discussed in previous sections suggest that the observed spectroscopic properties 
of GRB/CC SN sites will be largely affected by the blending with nearby HIIRs. 
In this section, we discuss what spatial resolution would be necessary 
to obtain a reliable metallicity for a GRB/CC SN site, 
and how the observed metallicities differ from the intrinsic ones 
under the observations with insufficient spatial resolution. 
In the following, we define the ``apparent metallicity" 
of an HIIR for a resolution scale radius of $R_{\rm res}$ as 
a metallicity inferred from the ratio of total H$\alpha$ and [\ion{N}{ii}] line fluxes of all HIIRs inside $R_{\rm res}$. 

As mentioned in \S\ref{sec:data}, the sampling rate of spectroscopic sample is low, 
and the N2 index is not available for most A11 HIIRs. 
For demonstration purposes, we assume 
that the HIIRs with unknown N2 index 
follows the same N2 index distribution as found in the S12 sample. 
We divide the S12 sample into four different bins of $R_{\rm deproj}$ 
(0--10, 10--15, 15--20, and $> 20$\,kpc, Figure~\ref{fig:metaldist}) 
and examine the effect of metallicity gradient. 

Practically, the apparent metallicities are calculated as follows. 
For each HIIR with H$\alpha$ and [\ion{N}{ii}] detections in S12, 
we collect all HIIRs from the A11 sample 
which reside within a given radius of $R_{\rm res}$ (deprojected), 
and sum up the H$\alpha$ and [\ion{N}{ii}] fluxes. 
For an A11 HIIR that has a matched S12 HIIR with a known N2 index, 
the H$\alpha$ and [\ion{N}{ii}] fluxes are calculated assuming the index value 
(see \S\ref{sec:lum} for the sample matching method). 
When the N2 index is not available for an A11 HIIR, 
we randomly select an S12 HIIR in the same $R_{\rm deproj}$ bin 
with H$\alpha$ and [\ion{N}{ii}] detections. 
Then we calculate the H$\alpha$ and [\ion{N}{ii}] fluxes of the A11 HIIR
assuming the same N2 index to that of the selected S12 HIIR.
We repeat this random realisation of apparent metallicities 100 times 
for each of 197 HIIRs with known N2 index and photometric counterparts. 

\begin{figure}
\includegraphics[width=84mm]{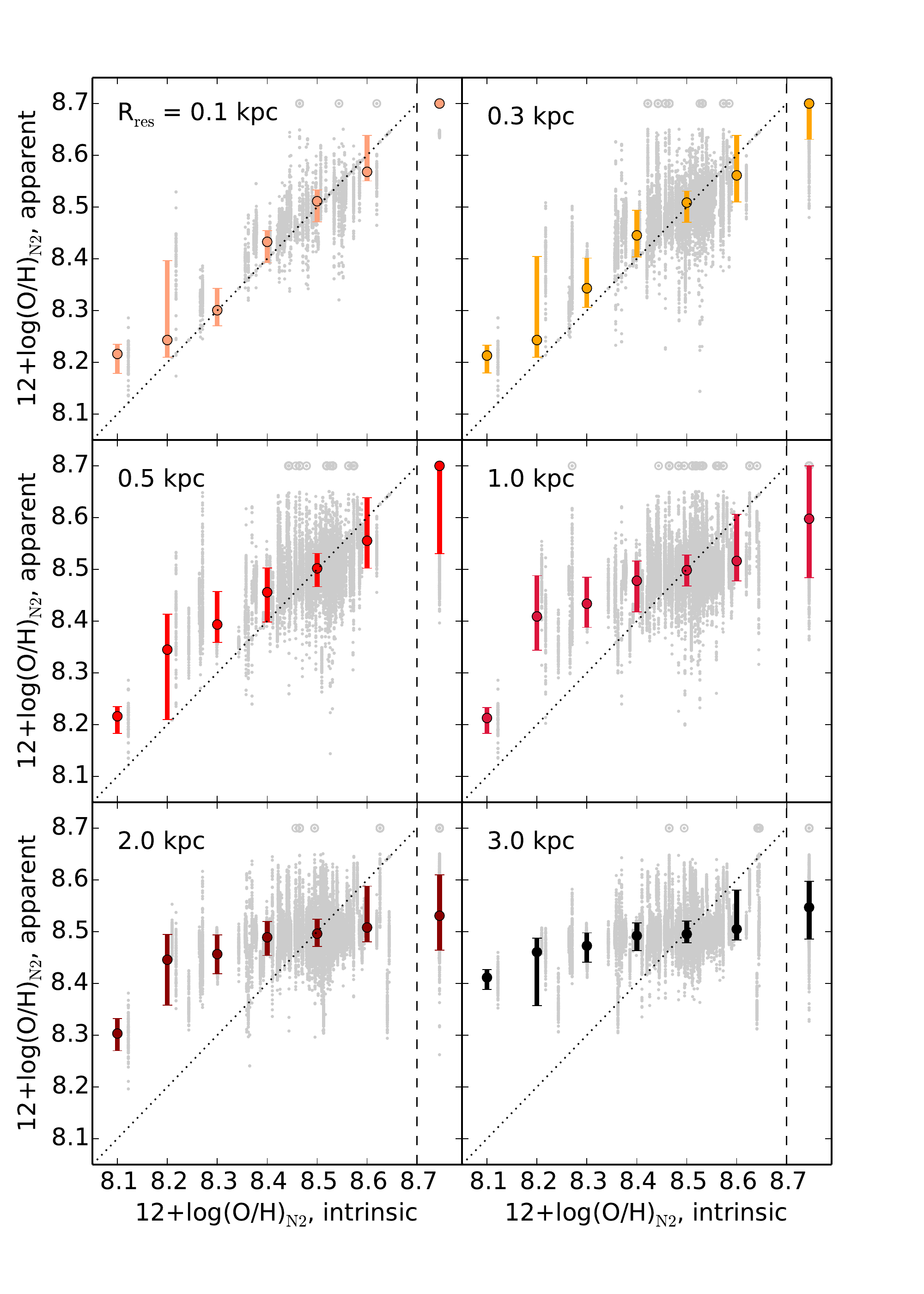}
\caption{
Intrinsic versus apparent metallicities of HIIRs 
with known N2 index for various $R_{\rm res}$. 
The grey dots represent each data points 
(197 HIIRs $\times$ 100 random realisations 
of apparent metallicity in each panel). 
The grey circles at the top of each panel 
represent the data with apparent N2 $> -0.2$. 
The data points with intrinsic N2 $> -0.2$ are plotted 
separately in the right-hand-side of each panel. 
The circles and the error bars indicate median and 68 percentile scatter 
of the data points in intrinsic metallicity bins (bin width of 0.2\,dex). 
The data points with apparent N2 $> -0.2$ 
are included in the median and scatter calculation. 
The diagonal dotted line indicates the equality between 
the apparent and intrinsic metallicities. 
}
\label{fig:ZintZapp}
\end{figure}

We show the intrinsic versus apparent metallicities of HIIRs
in Figure~\ref{fig:ZintZapp} for $R_{\rm res}$ = 0.1, 0.3, 0.5, 1.0, 2.0, and 3.0\,kpc. 
When $R_{\rm res} \geq 1$\,kpc, 
the HIIRs with lower (higher) intrinsic metallicities than 12+log(O/H)$_{\rm N2} = 8.5$ 
have systematically higher (lower) apparent metallicities than the intrinsic values. 
This is because most of the contaminating HIIRs
have N2 indices that correspond 
to 12+log(O/H) = 8.4--8.6 (Figure~\ref{fig:metaldist}). 
In this case, it is difficult to distinguish 
low-metallicity HIIRs from high-metallicity ones, 
because all HIIRs with intrinsic 12+log(O/H)$_{\rm N2}$ = 8.2--8.7 
have similar apparent metallicities within the scatter. 
With $R_{\rm res} \leq 0.5$\,kpc, the apparent metallicity 
correlates well with the intrinsic value, 
although some systematic differences 
exist for $R_{\rm res} = 0.5$\,kpc. 

As discussed in \S\ref{sec:lum}, 
the metallicities of low-luminosity HIIRs may be biased 
towards high metallicities (Figure~\ref{fig:LhaZ}). 
To investigate how the possible bias affects our results, 
we also compute the apparent metallicities 
considering the N2 index distribution of only bright HIIRs
with $\log_{10} L_{\rm H\alpha}  {\rm [erg\ s^{-1}]} > 36.5$ 
in the S12 sample when we randomly determine the N2 index 
of an A11 HIIR with unknown N2 index. 
We did not find significant differences between 
the results obtained with the N2 index distribution of bright HIIRs 
and those with all HIIRs, 
except that the apparent metallicities 
computed with the bright HIIRs 
are slightly lower (by $< 0.1$ dex) than those 
in Figure~\ref{fig:ZintZapp} for $R_{\rm res} \geq 1.0$\,kpc. 

\begin{figure}
\includegraphics[width=84mm]{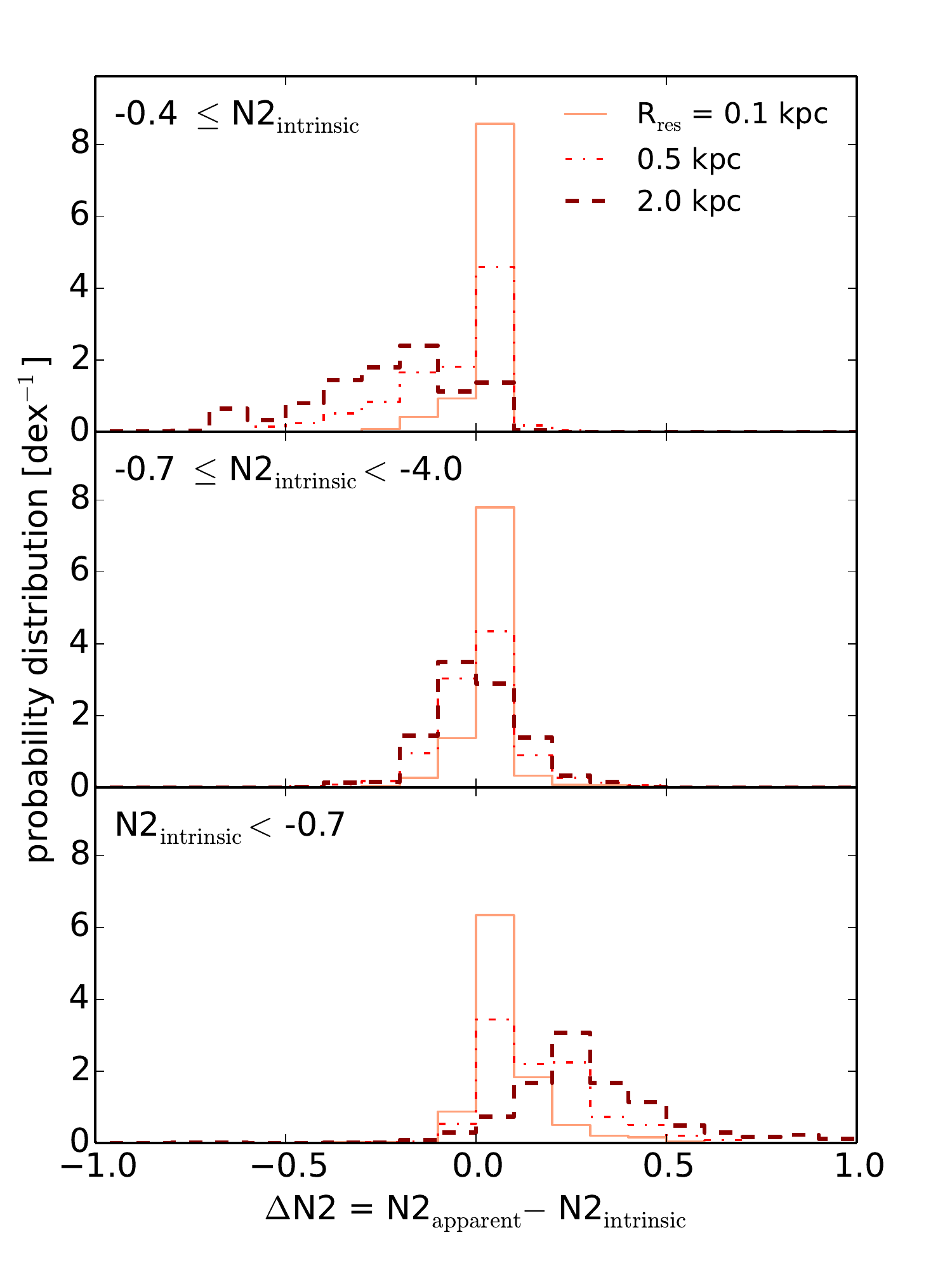}
\caption{
Probability distribution of difference between the intrinsic and apparent N2 index: 
 $\Delta$N2~=~N2$_{\rm apparent}-$~N2$_{\rm intrinsic}$. 
We divide the sample into the following three different ranges: 
N2 $< -0.7$ (bottom), $-0.7 \leq$ N2 $< -0.4$ (middle), and $-0.4 \leq$ N2 (top panel). 
}
\label{fig:dZ}
\end{figure}

In Figure~\ref{fig:dZ}, we quantify the deviation between the intrinsic and apparent 
metallicities in a different way from Figure~\ref{fig:ZintZapp}, 
by presenting the distributions of 
$\Delta$N2 = N2$_{\rm apparent}-$N2$_{\rm intrinsic}$ 
for low-, intermediate-, and high-metallicity HIIRs 
(i.e., intrinsic N2 index of $< -0.7$, $-0.7 \leq$ N2 $< -0.4$, and $-0.4 \leq$ N2, respectively) 
with $R_{\rm res} = 0.1$, 0.5, and 2.0\,kpc. 
For $R_{\rm res} = 0.1$\,kpc, 
the majority of HIIRs have $|\Delta$N2$|<0.1$ 
for all ranges of intrinsic metallicity. 
For $R_{\rm res} = 0.5$\,kpc, 
the $\Delta$N2 distribution still peaks at around zero, 
although there is a systematic difference 
between the N2$_{\rm intrinsic}$ and N2$_{\rm apparent}$
[i.e., $\Delta$N2 $>$ 0 for low-metallicity HIIRs and $<$ 0 for high-metallicity ones, 
which is consistent with what we see in Figure~\ref{fig:ZintZapp}].
For $R_{\rm res} = 2.0$\,kpc, there is no peak 
at $\Delta \log_{10}$(O/H) = 0 either for low- or high-metallicity HIIRs, 
meaning that the intrinsic and apparent metallicities do not agree with each other. 

We note that the metallicity variation in each HIIR is not considered in our analysis.  
Observations of some giant HIIRs in the Milky Way and the Magellanic clouds  
have shown that the metallicity variation within each HIIR is typically within $\pm0.1$ dex 
in most parts of the HIIRs \citep[e.g.,][]{Lebouteiller:2008a, Mesa-Delgado:2008a, Mesa-Delgado:2010a}. 
Thus we consider that the metallicity variation within each HIIR 
is smaller compared to that between different HIIRs. 
As mentioned in \S\ref{sec:diagnostic}, 
the N2 index and many other metallicity indicators 
are also affected by the physical conditions of gas other than metallicity. 
Hence it is unclear to what extent the observed variation 
of line ratios in M31 originates from the metallicity variation. 
However, we can at least conclude that there is a difficulty 
in measuring the intrinsic line ratio of a transient event site with $R_{\rm res} \geq 1$\,kpc. 

\section{Other Metallicity Indexes}
\label{sec:otherZ}

In this section, we discuss how the blending effect appears 
when we measure the HIIR metallicities with indicators other than N2 index. 
Here we consider, 
O3N2 index = log$_{10}$[([\ion{O}{iii}]~$\lambda$5007/H$\beta$)$\times$(H$\alpha$/[\ion{N}{ii}])], 
N2O2 index  = log$_{10}$([\ion{N}{ii}]/[\ion{O}{ii}]~$\lambda$3727), 
and $R_{23}$ index = ([\ion{O}{ii}]~$\lambda$3727+[\ion{O}{iii}]~$\lambda\lambda$4959, 5007)/H$\beta$, 
which are available with flux errors $< 20$ per cent for 58, 98, and 60 HIIRs, respectively. 
For the O3N2 indices, we use the metallicity calibration by \citet{Marino:2013a} 
based on the same HIIR sample as the N2 calibration. 
For O2N2 and $R_{23}$ indices, we use the theoretical calibration 
by \citet[][KD02]{Kewley:2002a} and \citet[][KK04]{Kobulnicky:2004a}, respectively. 

It is known that the $R_{23}$ index 
provides two solutions of possible metallicities, 
one in the upper branch [12+log(O/H)$_{R23} > 8.4$], 
and the other in the lower branch [12+log(O/H)$_{R23} < 8.4$]. 
KK04 showed that the HIIRs with 12+log(O/H)$_{R23} < 8.4$ hardly have N2 $> -1.0$, 
while more than 95 per cent of the S12 HIIR with N2 index have N2 $> -1.0$ 
[12+log(O/H)$_{\rm N2} >$ 8.28 in the calibration by \citet{Marino:2013a}, see Figure~\ref{fig:metaldist}]. 
Therefore, we assume the upper branch solutions for the M31 HIIRs,
although some HIIRs with N2 $< -1.0$ may have lower-branch metallicities. 

Similarly to the N2 index, 
other indices (O3N2, N2O2, and $R_{23}$) are calibrated to metallicity 
only in some limited ranges. 
In \citet{Marino:2013a}, the O3N2 index is calibrated  
within $-1.1<$ O3N2 $<1.7$ which corresponds 
to $8.17<$ 12+log(O/H) $_{\rm O3N2} < 8.77$, 
and the metallicity cannot be measured correctly with the O3N2 index out of this range. 
A few of the S12 HIIR have O3N2 $>1.7$, 
and we assume that these HIIRs have 12+log(O/H)$_{\rm O3N2} \lesssim 8.0$ 
(O3N2 index anti-correlates with metallicity). 
The N2O2 index is sensitive to metallicity 
when N2O2 $> -0.97$ [or 12+log(O/H)$_{\rm N2O2} > 8.6$, KD02], 
and we assume that the HIIRs with N2O2 $\leq -0.97$ have 12+log(O/H)$_{\rm N2O2} \lesssim 8.5$. 
 The $R_{23}$ index is calibrated at log$_{10}R_{23} \lesssim 1.0$ 
depending on an additional parameter $O_{32}$ = ([\ion{O}{iii}]~$\lambda\lambda$4959, 5007/[\ion{O}{ii}]~$\lambda$3727). 
All S12 HIIRs with $R_{23}$ index have the values of $R_{23}$ and $O_{32}$ 
in the calibrated ranges.

\begin{figure*}
\includegraphics[width=168mm]{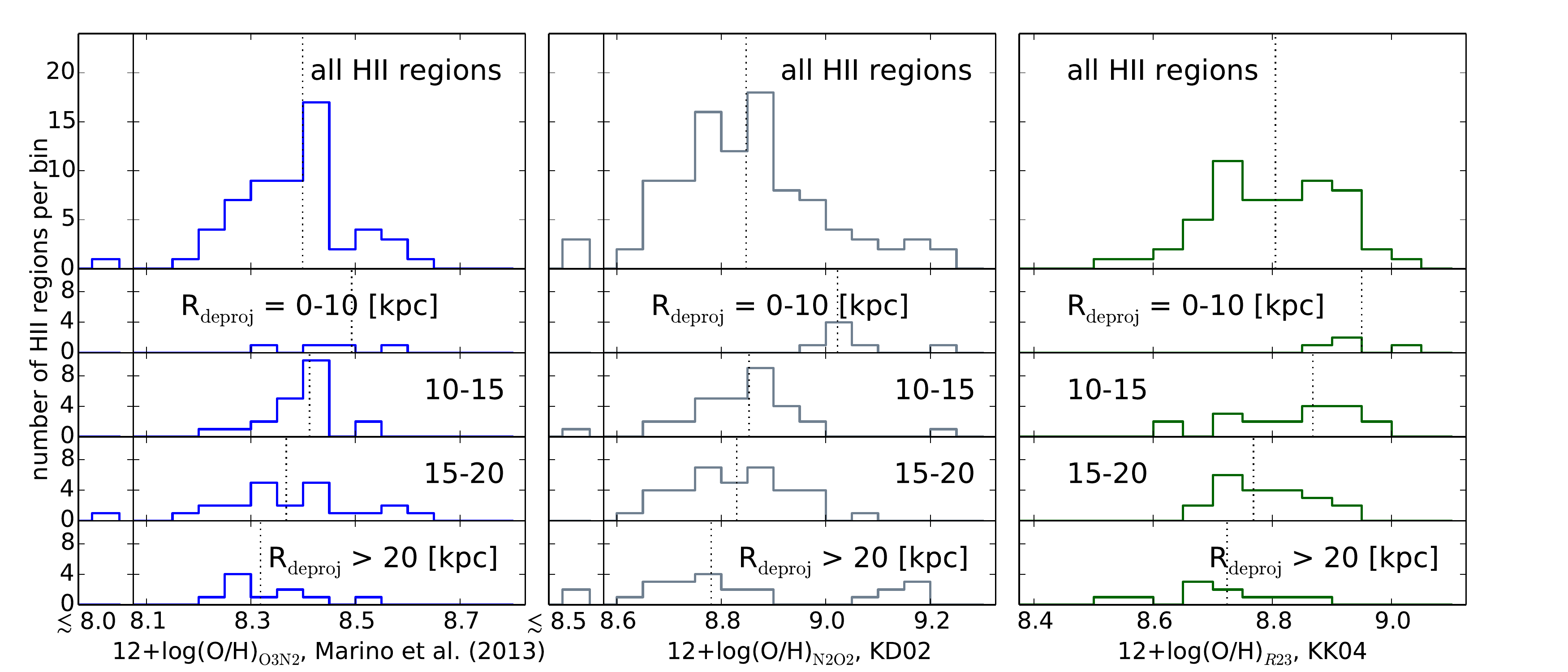}
\caption{
Same as Figure~\ref{fig:metaldist}, but for the metallicities 
measured with O3N2 (left), N2O2 (middle), and $R_{23}$ (right panel) indices 
using calibrations by \citet{Marino:2013a}, KD02, and KK04, respectively. 
}
\label{fig:metaldist_oth}
\end{figure*}

We show the metallicity distributions of M31 HIIRs measured 
with O3N2, N2O2, and $R_{23}$ indices in Figure~\ref{fig:metaldist_oth}. 
As previously reported, the theoretically calibrated methods 
give systematically higher metallicities than the empirical calibrations 
based on the `direct' method \citep{Kewley:2008a, Lopez-Sanchez:2010a}. 
The median 12+log(O/H) measured with N2, O3N2, N2O2, and $R_{23}$ indices 
are 8.51, 8.38, 8.85, and 8.81, respectively.  
All metallicity indices discussed here commonly 
have wide distributions of 12+log(O/H) spanning $\sim$ 0.6 dex with different sample sizes. 
The calibration of O3N2 index by \citet{Marino:2013a} has intrinsic errors 
of $\pm$0.08--0.18 dex depending on the HIIR sample used for the calibration, 
while \citet{Kewley:2008a} suggested that the calibrations 
by KD02 and KK04 have intrinsic error of 0.06 dex. 
The metallicity gradient is more significant with O3N2, N2O2, and $R_{23}$ indices than with N2 index, 
while the large scatter in each $R_{\rm deproj}$ is commonly seen. 
We note that the dependence of metallicity gradient 
on metallicity calibration method is also reported by S12. 
The N2O2 index possibly shows a multi-peak metallicity distribution 
at $R_{\rm deproj} >$ 20\,kpc which is not seen with other metallicity indices. 

\begin{figure*}
\includegraphics[width=168mm]{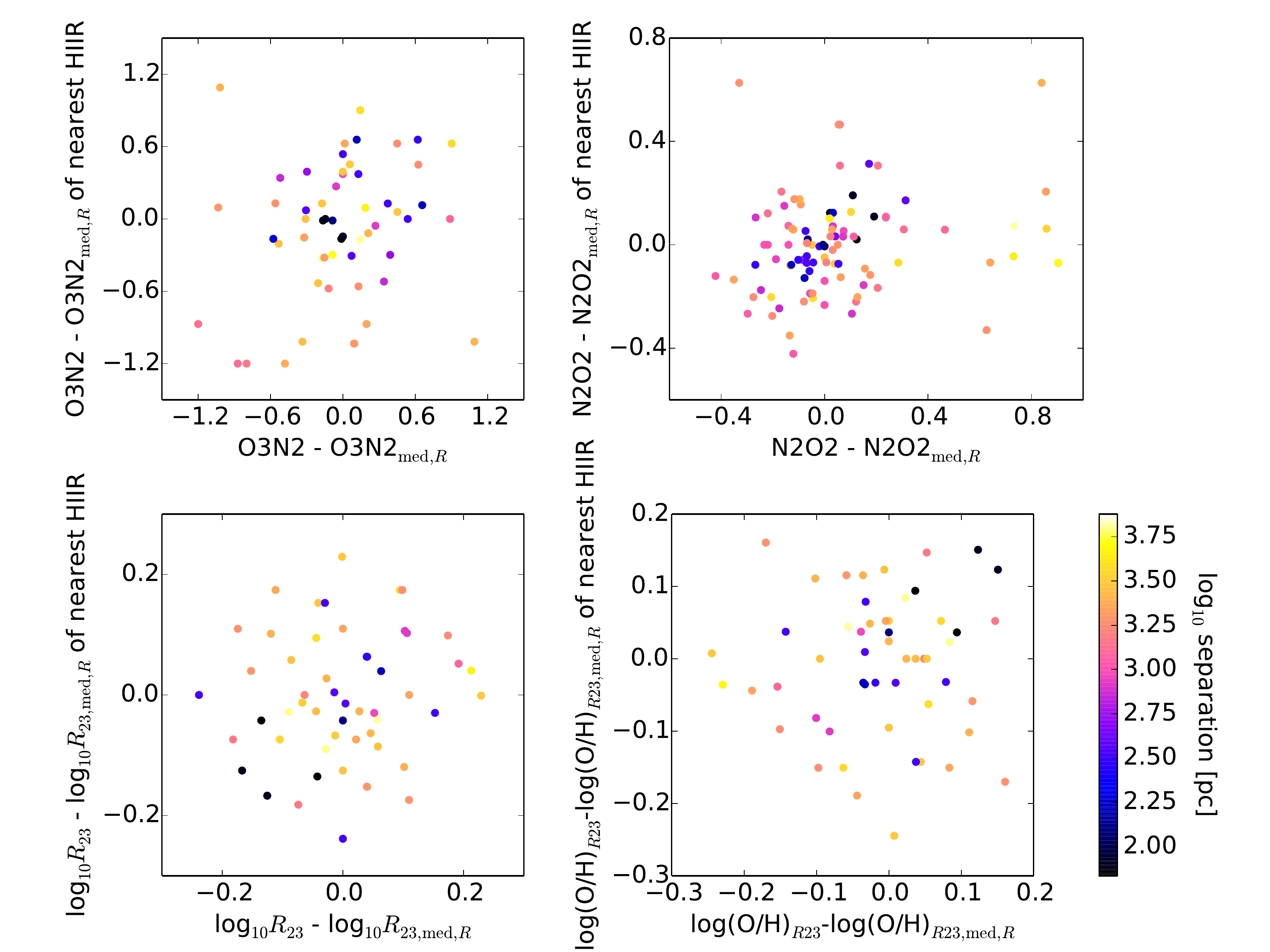}
\caption{
Same as Figure~\ref{fig:pairLR}, 
but for O3N2, N2O2, log$_{10}R_{23}$, and log(O/H)$_{R23}$ 
(top left, top right, bottom left, and bottom right panels, respectively). 
}
\label{fig:pairLR_oth}
\end{figure*}

Here we investigate the small-scale scatter of different metallicity indices 
similarly to \S\ref{sec:scatter}. 
The upper left, upper right, and lower left panels of Figure~\ref{fig:pairLR_oth} 
show the scatter of O3N2, N2O2, and $R_{23}$ indices, respectively. 
We discuss the scatter of the indices rather than the metallicity 
in order to include the HIIRs which have index values outside the calibrated range. 
While all three indices show large scatters 
similarly to the N2 index (Figure~\ref{fig:pairLR}), 
the N2O2 index also shows some correlations between 
the metallicities of neighboring HIIR pairs. 
The N2O2 index is not largely affected by the gas ionization state, 
however it suffers largely from the uncertainty of extinction correction (KD02). 
The calibration of $R_{23}$ index by KK04 also reduces 
the effect of ionization state using the additional parameter $O_{32}$, 
therefore the $R_{23}$ index and the metallicity do not correspond one-to-one.
In the lower right panel of Figure~\ref{fig:pairLR_oth}, 
we show the scatter of log(O/H) measured by the $R_{23}$ index (including $O_{32}$). 
With the $O_{32}$ parameter included, 
log(O/H)$_{R23}$ of the neighboring HIIR pairs show 
a large scatter as well as log$_{10}R_{23}$ itself. 
This suggests that the variation of ionization state is not the dominant source 
of the scatter seen in Figure~\ref{fig:pairLR}~and~\ref{fig:pairLR_oth}. 

\begin{figure}
\includegraphics[width=84mm]{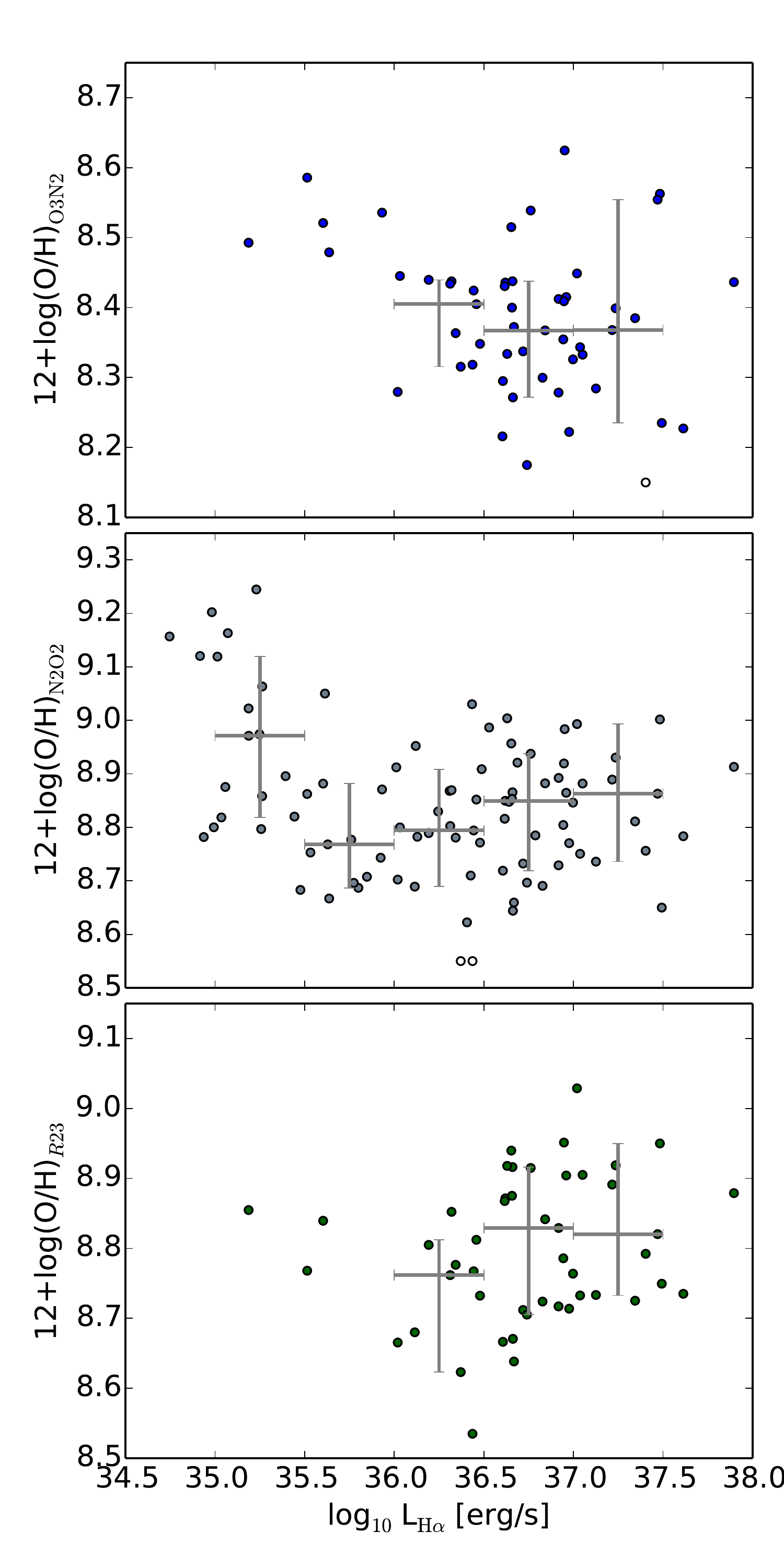}
\caption{
Same as Figure~\ref{fig:LhaZ}, 
but for the metallicities measured with O3N2, N2O2, and $R_{23}$ indices 
(top, middle, and bottom panels, respectively). 
}
\label{fig:LhaZ_oth}
\end{figure}

In Figure~\ref{fig:LhaZ_oth}, we plot $L_{\rm H\alpha}$ 
versus metallicity of HIIRs measured with O3N2, N2O2, and $R_{23}$ indices. 
The O3N2 and $R_{23}$ indices (the top and bottom panels, respectively) 
are rarely available for faint HIIRs with log$_{10} L_{\rm H\alpha}$ [erg s$^{-1}$] $< 36.0$. 
However, a small number of faint HIIRs with O3N2 index 
have systematically higher 12+log(O/H)$_{\rm O3N2}$ 
than the HIIRs with larger $L_{\rm H\alpha}$, 
which is consistent with the trend found with the N2 index. 
On the other hand, such a trend is not seen with $R_{23}$. 
With the N2O2 index, the metallicity is constant over a wide range 
of H$\alpha$ luminosity, $35.5 <$ log$_{10} L_{\rm H\alpha}$ [erg s$^{-1}$] $< 37.5$ 
(or possibly higher metallicity with larger $L_{\rm H\alpha}$), 
although the faintest HIIRs with log$_{10} L_{\rm H\alpha}$ [erg s$^{-1}$] $\sim 35$ 
have systematically higher metallicities than those with larger $L_{\rm H\alpha}$. 
The dependence of $L_{\rm H\alpha}$--metallicity relation 
on the metallicity indicator suggests that the relation 
is produced by some artificial effect, 
although it is difficult to draw robust conclusions. 

\begin{figure*}
\includegraphics[width=147mm]{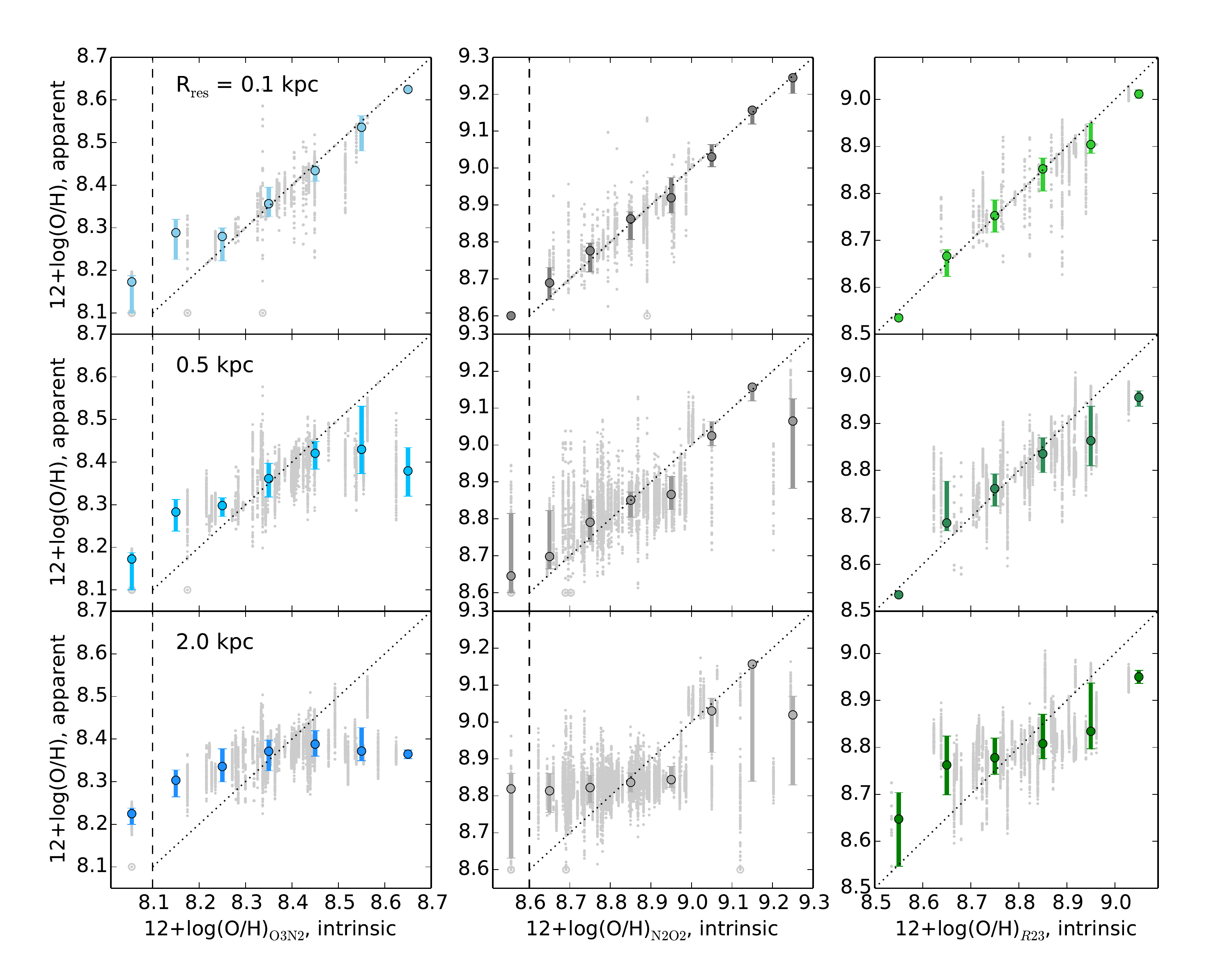}
\caption{
Same as Figure~\ref{fig:ZintZapp}, 
but for the metallicities measured with the O3N2, N2O2, and $R_{23}$ indices 
(left, middle, and right columns of panels, respectively). 
Results for $R_{\rm res}$ = 0.1, 0.5, and 2.0\,kpc 
are shown in the top, middle, and bottom panels, respectively. 
}
\label{fig:ZintZapp_oth}
\end{figure*}

In Figure~\ref{fig:ZintZapp_oth},  
we show the relation between the intrinsic and apparent metallicities 
using the O3N2, N2O2, and $R_{23}$ indices 
with $R_{\rm res}$ = 0.1, 0.5, and 2.0\,kpc 
The apparent metallicities are computed following the same method 
as for the N2 index (\S\ref{sec:demo}). 
Although the absolute metallicity scale is different for different metallicity indices, 
the relations between the intrinsic and apparent metallicities 
are more or less consistent with the N2 index case. 

When the N2O2 index is used, the HIIRs with intrinsic 12+log(O/H) $> 9.0$ 
show a bimodal distribution of apparent metallicities 
[12+log(O/H) $\sim 8.8$ and 9.0] for $R_{\rm res}$ = 0.5, and 2.0\,kpc. 
This is due to the possible multi-peak metallicity distribution of HIIRs 
at $R_{\rm deproj} >$ 20\,kpc which appears only when the N2O2 index is used 
(the middle panel Figure~\ref{fig:metaldist_oth}). 
The HIIRs with intrinsic 12+log(O/H)$_{\rm N2O2} > 9.0$ mainly reside 
at $R_{\rm deproj} =$ 0--10\,kpc, and $>$ 20\,kpc. 
When $R_{\rm res}$ is not small enough to measure the intrinsic metallicities, 
the HIIRs at $R_{\rm deproj} =$ 0--10\,kpc have apparent 12+log(O/H)$_{\rm N2O2} \sim 9.0$, 
while those at $R_{\rm deproj} >$ 20\,kpc have apparent 12+log(O/H)$_{\rm N2O2} \sim 8.8$, 
which are the typical metallicities in those $R_{\rm deproj}$ bins. 

\section{Implications for GRB/CC SN Site Studies}
\label{sec:discussion}

Many CC SNe are found in the local Universe ($\lesssim$ 100\,Mpc), 
where typical angular resolution of a ground-based optical observation
($\sim$ 1 arcsec) corresponds to $\lesssim 500$ pc. 
The metallicities of CC SN sites in the local Universe are usually investigated 
with spatial resolution of less than a few hundred pc 
\citep[e.g.,][]{Anderson:2010a, Sanders:2012b, Kuncarayakti:2013a}, 
and our results suggest that the metallicities measured with such spatial resolution 
can be used as a proxy to study the immediate environment of transient events, 
if ISM properties of their host galaxies are similar to those in M31. 
However, there might be systematic differences between the apparent and intrinsic metallicities 
when the spatial resolution exceeds a few hundred pc. 

\begin{table*}
 \centering
 \begin{minipage}{140mm}
  \caption{Emission lines observed at long GRB sites in the literature. 
    The line fluxes are from GRB 980425: \citet{Christensen:2008a}, 020819: \citet{Levesque:2010b}, 
    060505: \citet{Thone:2008a}, \citet{Thone:2014a}, 100316D: \citet{Levesque:2011a}, 
    120422A: \citet{Levesque:2012a}, and \citet{Schulze:2014a}, in units of $10^{-16}$ erg s$^{-1}$ cm$^{-2}$. 
    The flux errors are typically $\sim$ 10 per cent unless otherwise specified. 
  }
  \label{tab:siteSpec}
  \begin{tabular}{@{}lrrrrrrrr@{}}
  \hline
   GRB & redshift & spatial & [\ion{O}{ii}]~$\lambda$3727 & H$\beta$ & [\ion{O}{iii}]~$\lambda$4959 & [\ion{O}{iii}]~$\lambda$5007 & H$\alpha$ & [\ion{N}{ii}] \\
    &  & resolution &  &  &  &  & \\
    &  & [kpc]\footnote{Diameter of seeing size unless otherwise specified.} &  &  &  &  &  &  \\
  \hline
   980425 & 0.0085 & 0.27\footnote{Diameter of the integral field unit resolution element \citep{Christensen:2008a}.} 
                                           & 26.07 & 7.55 & - & 18.27 & 44.94 & 5.03 \\
   020819 & 0.41 & 3.0 & 0.92 & - & - & - & 4.14 & 1.51 \\
   060505\footnote{Two independent sets of emission line fluxes are obtained 
        for the site of GRB 060505 by \citet[][upper row]{Thone:2008a} \& \citet[][lower row]{Thone:2014a}} 
        & 0.089 & 1.2 & 1.12 & 0.298 & 0.284 & 0.72 & 0.973 & 0.096 \\
    &  & 2.2 & 2.39$\pm$0.13 & 1.15$\pm$0.07 & 0.76$\pm$0.08 & 1.98$\pm$0.06 & 3.11$\pm$0.07 & 0.24$\pm$0.03 \\
   100316D & 0.059 & 1.0 & - & 90$\pm$10 & - & 280$\pm$10 & 200$\pm$10 & 12$\pm$10 \\
   120422A\footnote{Two independent sets of emission line fluxes are obtained 
        for the site of GRB 120422A by \citet[][upper row]{Levesque:2012a} \& \citet[][lower row]{Schulze:2014a}} 
        & 0.283 & 3.0 & 0.99$\pm$0.07 & 0.16$\pm$0.01 & $< 0.11$ & 0.42$\pm$0.03 & 0.62$\pm$0.04 & $< 0.08$ \\
    &  & 4.2 & 0.25$\pm$0.01 & 0.05$\pm$0.04 & 0.05$\pm$0.02 & 0.19$\pm$0.02 & 0.24$\pm$0.01 & 0.06$\pm$0.02 \\
  \hline
\end{tabular}
\end{minipage}
\end{table*}

When we study the transient events 
whose rate density is much lower than that of CC SNe 
(e.g., long GRBs, $\sim$ several $\times 10^{-7}$ yr$^{-1}$Mpc$^{-3}$), 
chances of detecting an event in the local Universe is small, 
and we need to rely on samples at larger distances. 
For now, spatially resolved spectra of long GRB sites 
are obtained for 5 low-redshift GRBs. 
In Table~\ref{tab:siteSpec}, we summarise the redshifts 
and relevant emission line fluxes of the sites of these bursts, 
together with the spatial resolution of each observation. 
The fluxes are corrected for the extinction in the Milky Way, but not for the host galaxy. 
The sites of GRB 060505 and 120422A have 
two sets of emission line fluxes obtained independently. 
Although the emission line fluxes of the site of GRB 120422A 
corrected for the host galaxy extinction are provided in \citet{Levesque:2012a}, 
we undo this correction for consistency with other data. 

In Table~\ref{tab:siteZs}, we show $E_{B-V}$ in the host galaxy, 
and the site metallicities measured with the N2, O3N2, N2O2, and $R_{23}$ indices 
computed from the line fluxes listed in Table~\ref{tab:siteSpec}. 
The value of $E_{B-V}$ is obtained using H$\alpha$/H$\beta$ line ratio
assuming the intrinsic value of H$\alpha$/H$\beta$ = 2.85 and the Cardelli extinction curve. 
The metallicity indices are extinction corrected accordingly. 
When the index values are outside of the calibrated range, 
we show the index value instead of the metallicities.
It is clear that the different metallicity calibration methods 
give different metallicities for the same GRB sites, 
and thus we compare the metallicities of the GRB sites 
only when they are obtained by the same calibration method. 

The explosion site of GRB 980425 is the only one 
which is spectroscopically observed with a comparable spatial resolution 
to $R_{\rm res} \sim$ 0.1\,kpc owing to its close distance of $z = 0.0085$ \citep{Christensen:2008a}. 
With the N2 index, the explosion site metallicity of this burst 
is 12+log(O/H)$_{\rm N2}$ = 8.30 with the calibration by \citet{Marino:2013a}. 
In the other cases, the spatial resolution is $\gtrsim$\,1\,kpc, 
and our results suggest that there might be systematic differences 
between the observed and the actual metallicity. 

The sites of GRB 020819 and 120422A \citep{Levesque:2010b, Schulze:2014a} 
have higher metallicities than other explosion sites when the N2 index is used 
[12+log(O/H)$_{\rm N2} \sim 8.5$ which is similar to the typical metallicity of M31 HIIRs], 
although the independent spectroscopy of the GRB 120422A 
site by \citet{Levesque:2012a} indicates a lower metallicity 
with the same metallicity calibration method 
but with different slit alignment and seeing size 
(N2 $< -0.89$ by \citeauthor{Levesque:2012a}~\citeyear{Levesque:2012a}, 
N2 $= -0.60^{+0.12}_{-0.18}$ by \citeauthor{Schulze:2014a}~\citeyear{Schulze:2014a}). 
The site of GRB 060505 \citep{Thone:2014a} also shows a high metallicity when the $R_{23}$ index is used, 
although the metallicity of this site is similar to the other sites when measured with other indices. 
A lower metallicity is independently suggested also at the GRB 060505 site 
using the emission line fluxes obtained by \citet{Thone:2008a}. 
As mentioned in \S\ref{sec:otherZ}, the calibration methods 
using the $R_{23}$ index provides two possible solutions of metallicities. 
According to the photoionization model of KK04, 
N2 = $-1.1$--$-1.0$ and $O_{32}$ = 0.9--1.1 obtained at the GRB 060505 site 
suggests that the upper branch solution [12+log(O/H)$_{R23} > 8.4$] is the correct one. 

The GRB sites with high metallicities 
are especially interesting in the context of GRB progenitor studies, 
casting doubts on the low metallicities of GRB progenitors predicted 
by some theoretical studies \citep[e.g.,][]{Yoon:2005a,Woosley:2006a}. 
However,  
the high-metallicity GRB sites are currently found 
only by the spectroscopic observations with spatial resolution of $> 2$\,kpc. 
In our analysis, these spatial resolution correspond to the cases with $R_{\rm res} \geq 1.0$\,kpc, 
where one cannot distinguish low and high metallicities. 
Thus the observed high metallicities could be significantly different 
from the true values in the immediate environment of the GRBs. 

We note that the metallicity of the GRB host galaxies 
are also studied via absorption lines in the GRB afterglow spectra
\citep[e.g.,][]{Fynbo:2008a, Savaglio:2009a, Cucchiara:2014a}, 
and high metallicities are found in some cases \citep[e.g.,][]{Savaglio:2012a}. 
However, the absorption line study would give the metallicity averaged 
over the line-of-sight in the host galaxy, and its relation with the metallicity 
in the immediate environment of a GRB is not understood well. 

\begin{table*}
 \centering
 \begin{minipage}{140mm}
  \caption{
    Extinctions and metallicities of the long GRB sites 
    computed from the emission-line fluxes listed in Table~\ref{tab:siteSpec}. 
    For the sites of GRB 020819 \citep{Levesque:2010b} 
    and GRB 120422A \citep[as observed by][]{Schulze:2014a}, 
    we only use the N2 index to measure the metallicities 
    due to the non-detection or the large flux errors in H$\beta$, 
    and the resulting poor constraint on $E_{B-V}$.
  } 
  \label{tab:siteZs}
  \begin{tabular}{@{}lrrrrr@{}}
  \hline
   GRB & $E_{B-V}$ & 12+log(O/H)$_{\rm N2}$ & 12+log(O/H)$_{\rm O3N2}$ & 12+log(O/H)$_{\rm N2O2}$ & 12+log(O/H)$_{R23}$ \\
  \hline
   980425 & 0.74$\pm$0.13 & 8.30 & 8.26 & (N2O2 = $-1.39$) & (log$_{10}R_{23}=0.99$, \\
    &  &  &  &  & log$_{10}O_{32}=-0.54$)  \\
   020819 & - & 8.54 & - & - & - \\
   060505 & 0.13$\pm$0.13 & 8.28 & 8.24 & (N2O2 = $-1.3$--$-1.1$) & 8.51 \\
    & 0.0 & 8.23 & 8.24 & (N2O2 = $-1.0$) & 8.82 \\
   100316D & 0.0 & 8.18 & 8.17 & - & - \\
   120422A & 0.31$\pm$0.09 & $<$ 8.33 & $<$ 8.26 & (N2O2 $< -1.37$) & (log$_{10}R_{23}=1.04$, \\
    &  &  &  &  & log$_{10}O_{32}=-0.53$)  \\
    & 0.0--1.1 & 8.47 & - & - & - \\
  \hline
\end{tabular}
\end{minipage}
\end{table*}

\section{Conclusions}
\label{sec:fin}

In this paper we examine how the small-scale metallicity variation in a galaxy 
affects the observations of GRB/CC SN sites with limited spatial resolution, 
using the observational data of M31 HIIRs
as a template of metallicity variation in a late-type galaxy. 
Our results suggest that, 
when the GRB/CC SN sites are resolved down to $R_{\rm res}  \lesssim 500$ pc scale, 
the estimated apparent metallicities (or emission-line ratios) 
do reflect the immediate environment of the transient events, 
but with significant systematic errors. 
If $R_{\rm res} \lesssim 100$ pc, the measured metallicity 
correlates with the intrinsic one tightly. 

The CC SN site studies are often conducted in the local Universe ($< 100$ Mpc),  
and we can achieve $R_{\rm res} \lesssim$ a few hundred pc. 
On the other hand, the detections of transient events with low event rate density 
(e.g., long GRBs) are rare in the local Universe, 
and we need to rely on a sample at higher redshifts,  
where $R_{\rm res} \gtrsim$ a few kpc  typically. 
With such low spatial resolution, our results suggest 
that it is difficult to constrain the site metallicities accurately,  
and the measured HIIR metallicities will be 
close to the average metallicity of the host galaxy 
due to the blending within the spatial resolution. 

However, we would like to emphasize that the overall properties of host galaxies 
are still important clues to the nature of transient progenitors. 
For example, the transient events that originate from low-metallicity stars 
would occur preferentially (but not exclusively) in low-metallicity galaxies. 

Some of the host galaxies of long GRBs and the explosion sites 
are found to have high metallicities 
with limited spatial resolution of $\gtrsim$ a few kpc
\citep[e.g.,][]{Levesque:2010b, Niino:2012b, Elliott:2013a, Schulze:2014a, Hashimoto:2014a}. 
However, low-metallicity star formation could still take place 
in a host galaxy with a high averaged metallicity. 
In fact, \citet{Niino:2011b} showed that up to $\sim$ 25 per cent 
of cosmic low-metallicity star formation in the local Universe 
takes place in high-metallicity galaxies with 12+log(O/H) $>$ 8.8. 
This means that 
{\em 
the above observations do not rule out the hypothesis 
that long GRBs are exclusively born in a low-metallicity environment, 
as suggested by the stellar evolution models 
} 
\citep[e.g.,][]{Yoon:2005a,Woosley:2006a}.

Our results are based only on the observed statistical properties of the HIIRs in M31, 
however, HIIRs in other galaxies 
may have different properties to the M31 HIIRs. 
Many GRB host galaxies are in fact dwarf irregulars that actively form stars \citep[e.g.,][]{Fruchter:2006a}, 
and it is likely that they have different ISM properties to that of the spiral galaxies like M31. 
At the same time, some host galaxies of long GRBs are spiral galaxies, 
which often have higher masses and metallicities than dwarf irregulars. 
These spiral galaxies are especially interesting in the context 
of metallicity dependence of long-GRB occurrence, and they may have 
similar HIIR properties to M31. 
Larger spectroscopic samples of HIIRs 
with very high spatial resolution of a few 10 pc, 
as recently done by \citet{Sanchez:2014b}, 
in dwarf irregulars and other types of galaxies
are necessary to further discuss the relation between the apparent and intrinsic metallicities 
in more general population of host galaxies. 

With a seeing size of $\sim$\,1\,arcsec, which is typical of ground-based optical observations, 
we can achieve the spatial resolution of $R_{\rm res} < 1$\,kpc only at redshifts $< 0.1$. 
To investigate the immediate environments of transients at $z \gtrsim 0.1$ 
with sufficient spatial resolution, we need observations with space telescopes. 
Observations in different wavelengths other than the optical 
may also be a solution once good metallicity diagnostics are found, 
although currently there are no known metallicity diagnostics 
in the wavelength ranges where we can achieve high spatial resolution from the ground
(e.g., in near-infrared with adaptive optics or in radio with interferometers) 
unless they are significantly redshifted \citep[e.g.,][]{Giveon:2002a, Nagao:2012a}.  

\section*{Acknowledgments}

We would like to thank our referee, \'Angel R, L\'opez-S\'anchez, for his helpful comments. 
We also thank Antonino Cucchiara, Raffaella Margutti, Maryam Modjaz, 
Nathan Sanders, and Francesco Taddia for their useful comments. 
Y.N. is supported by the Research Fellowship for Young Scientists 
from the Japan Society for the Promotion of Science (JSPS). 

\bibliographystyle{mn2e}
\bibliography{reference_list}

\bsp

\label{lastpage}

\end{document}